\documentclass[oldversion]{aa}  
\usepackage{graphicx,amsmath}
\newcommand{\bpicb}{$\beta$\,Pic\,b}
\newcommand{\bp}{$\beta$\,Pic}

\begin{document}

\title{The orbit of Beta\,Pic\,b as a transiting planet}

\author{
A.~Lecavelier des Etangs\inst{1,2}
\and
A.~Vidal-Madjar\inst{1,2}       
}
   
%\authorrunning{A.~Lecavelier \& A.~Vidal-Madjar}
%\titlerunning{The orbit of $\beta$\,Pic\,b as a transiting planet}

\offprints{A.L. (\email{lecaveli@iap.fr})}

\institute{
CNRS, UMR 7095, 
Institut d'Astrophysique de Paris, 
98$^{\rm bis}$ boulevard Arago, F-75014 Paris, France
   \and
UPMC Univ. Paris 6, UMR 7095, 
Institut d'Astrophysique de Paris, 
98$^{\rm bis}$ boulevard Arago, F-75014 Paris, France
}
   
\date{} %Received ...; accepted ...}
 
\abstract
{
In 1981, $\beta$\,Pictoris showed strong and rapid photometric variations possibly 
due to a transiting giant planet. 
Later, a planetary mass companion to the star, \bpicb, was identified using imagery. 
Observations at different epochs (2003 and 2009-2015) detected the planet 
at a projected distance of 6 to 9\,AU from the star and showed that the planet 
is on an edge-on orbit. The observed motion is consistent with an inferior conjunction 
in 1981, and \bpicb\ can be the transiting planet proposed to explain the 
photometric event observed at that time. Assuming that the 1981 event is related to 
the transit or the inferior conjunction of \bpicb\ on an edge-on orbit, 
we search for the planetary orbit in agreement with all the measurements 
of the planet position published so far. We find two different orbits that are 
compatible with all these constraints: (i) an orbit with a period of 17.97$\pm$0.08~years 
along with an eccentricity of around 0.12 and (ii) an orbit with a period 
of 36.38$\pm$0.13~years and a larger eccentricity of about 0.32. In the near future, 
new imaging observations should allow us to discriminate between these two different orbits. 
We also estimate the possible dates for the next transits, which could take place 
as early as 2017 or 2018, even for a long-period orbit. 
}


\maketitle
%
%________________________________________________________________

\section{Introduction}
\label{Introduction}

When its IR excess was detected by the IRAS satellite (Auman et al.\ 1983), 
\bp\ became the first star to be imaged with a circumstellar disk seen edge-on (Smith and Terrile, 1984). 
This particular disk is in fact a debris disk in the last stage of planetary formation.
This disk presents a wide variety of phenomena and components, including a dust disk, a gas disk, 
and falling and orbiting evaporating bodies (see review in Vidal-Madjar et al.\ 1998 where
the evidence for the presence of one or several planets is also discussed). 
The study of such a close-by and young planetary system (about 2$\times$10$^7$ years old; 
Binks et al.\ 2014; Malo et al.\ 2014) is of extreme interest because it is supposed 
to be in time after the formation of giant planets, but still possibly in the phase 
of on-going formation of satellites and telluric planets (Lagrange et al.\ 2000).

To explain the presence of numerous falling and evaporating bodies in the \bp\ system 
(FEBs, or exocomets), Beust et al. (1991) showed that a putative giant planet could be 
responsible for these bodies, if the planet has an orbit with an eccentricity 
of e~$\sim$~0.6 or more. 
Alternatively, the eccentricity-pumping effect of mean-motion resonances with 
a massive planet on a moderately eccentric orbit can also explain the orientation the exocomets' orbits. 
In particular, Beust \& Morbidelli (1996) show that the 4:1 mean-motion resonance 
is a very efficient mechanism for producting the evaporating exocomets 
as soon as the eccentricity of the perturbing planet is higher than about 0.05.
This last scenario is strengthened by the discovery of two families of exocomets, 
one of which presents the periastron angle-distance relationship for bodies trapped 
in a mean motion resonance (Kiefer et al.\ 2014). 

In this context, the discovery of a planet by direct imaging (Lagrange et al.\ 2009a) 
brings new and important information that enlightens what is happening 
in this young planetary system. 
The first series of images obtained by Lagrange et al.\ (2009a, 2009b, 2010) 
and Chauvin et al.\ (2012), shows that the planet has been directly detected on both sides of the disk, 
thus confirming the existence of the planet \bpicb\ with an orbital motion 
in the same plane as the circumstellar disk. 

The star \bp\ showed large photometric variations in November 1981 
(Lecavelier des Etangs et al.\ 1994, 1995). 
These variations were attributed to the transit of a planet orbiting at several AUs 
(Lecavelier des Etangs et al.\ 1994, 1995, 1997) or to a giant comet (Lamers et al.\ 1997). 
From analysis of the light curve and assuming that the variations are due to a transiting planet, 
Lecavelier des Etangs et al.\ (1997) obtained the following constraints: 
(1) if the planet is on a circular orbit, its period must be less than about 19~years 
(constrained by the measurements made on November 10 and 11, 1981) and 
(2) the size of the transiting object must be about 2 to 4 times the radius of Jupiter 
(constrained by the transit ingress in the light curve). 
The stellar limb-darkening effect was also detected during that transit event. 
A slight color effect was detected with more absorption at the shortest wavelengths in the U~band, 
which may be explained by dust particles around the occulting planet with Rayleigh scattering, 
as observed in the atmospheres of exoplanets (Lecavelier des Etangs et al.\ 2008a, 2008b). 
In all cases, material in the planet environment is needed to explain the large occultation depth 
that is too high to correspond to the normal size of a planet alone, 
even for a single hydrogen-dominated warm and inflated gaseous 
planet in a young system. The occultation depth is consistent with 
a circumplanetary (proto-satellite) dust disk or a ring system around the planet, 
as suggested for the planet Fomalhaut\,b in a similar young debris disk (Kalas et al.\ 2008).
Photometric surveys of \bpicb\ to search for another similar 
transit event gave only negative results. 
This showed that all short periods below 1~year and most periods below 2 to 3~years 
are excluded (Nitschelm et al.\ 2000; Lecavelier des Etangs et al.\ 2005).

In summary, we have the following detections: a planet orbiting within the edge-on disk of \bp\ 
and an object that transited \bp\ in 1981. 
The present paper further investigates the hypothesis that \bpicb\ is 
the transiting planet that was responsible for the photometric event recorded in November 1981. 

\section{New data and the transit scenario}

\subsection{Star-planet distance}

We now have a large set of measurements of the planet's position obtained with various telescopes and instruments.
In this work, we consider all the available measurements published by Currie et al.\ (2011), Chauvin et al.\ (2012), 
Absil et al.\ (2013), Bonnefoy et al.\ (2013), Males et al.\ (2014), Morzinski et al.\ (2014), 
Macintosh et al.\ (2014), Bonnefoy et al.\ (2014), Nielsen et al.\ (2014), and Millar-Blanchaer et al.\ (2015). 
This provides a set of 36 measurements for the star-planet distance 
and for the position angle (PA) of the star-planet direction relative to the north on the sky. 

\begin{figure}[tb]
\includegraphics[angle=90,width=\columnwidth]
{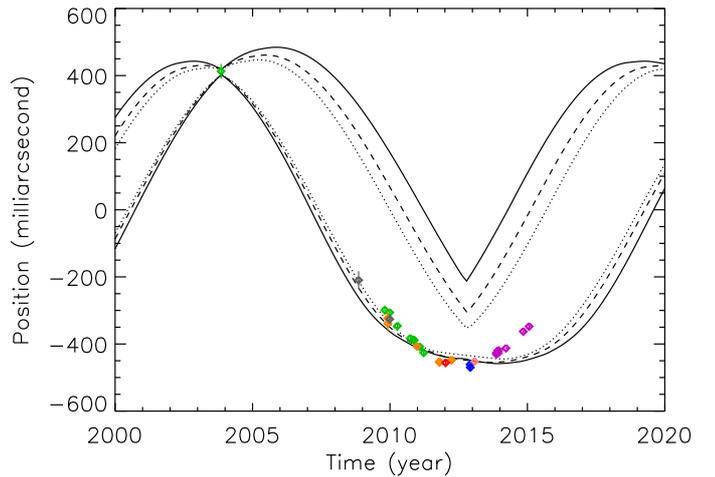}
\caption[]{
 Measured distance between \bp\ and \bpicb\ as a function of time (colored symbols), superimposed 
on the exact reproduction of the prediction given in Fig.~7 of 
Lecavelier des Etangs \& Vidal-Madjar (2009). The prediction 
was obtained using only the position of the planet measured in 2003 and the hypothesis of a transit 
in November 1981. Vertical bars represent 1-$\sigma$ error bars. 
The positive values correspond to the northeast branch of the disk where \bpicb\ was observed in 2003. 
The envelopes of the prediction, as published in the 2009 paper, are given for eccentricities 
of $e=$0.02 (dotted lines), $e=$0.05 (dashed lines), and $e=$0.1 (solid line). 
The measurements are taken from 
Currie et al.\ (2011, gray), Chauvin et al.\ (2012, green), 
Absil et al.\ (2013, pink), Bonnefoy et al.\ (2013, red),
Males et al.\ (2014, blue), Nielsen et al.\ (2014, orange), 
and Millar-Blanchaer et al.\ (2015, purple). 
\label{Position_Envelop_Fig7}}
\end{figure}

Most importantly, all the published measurements of the star-planet distance 
are in agreement with the predictions given by Lecavelier des Etangs \& Vidal-Madjar (2009). 
This prediction was obtained in 2009 using only two data sets to calculate the orbital motion 
of the planet: the single position measured in images made in 2003 
and the assumption that the planet was transiting on November 10, 1981. 
If we overplot the positions measured over 6~years between 2009 and 2015 to the reproduction of 
Fig.~7 of Lecavelier des Etangs \& Vidal-Madjar (2009), the new measurements well follow 
the envelope predicted for an eccentricity lower than 0.1 (Fig.~\ref{Position_Envelop_Fig7}).
This shows that the measurements of the star-planet distance gathered for 6~years
are in agreement with the idea that \bpicb\ is a transiting planet 
that was responsible for the photometric event recorded in November 1981.

\subsection{Does \bpicb\ transit? \\ 
Orbit inclination and the position angle of \bpicb}
\label{Orbit inclination}

The planet $\beta$\,Pic b is a transiting planet only if the inclination of its orbit 
differs from 90$\degr$ by less than $\pm 0.05\degr$. 
However, Millar-Blanchaer et al.\ (2015, hereafter MB15) 
showed that their last astrometric measurements constrain the inclination 
to be 89.0$\degr$$\pm$0.3$\degr$, i.e., 3-$\sigma$ away from a transit configuration.
This result can be understood by looking at the published measurements 
of the position angle (PA) of the planet. 
The variations of the PA values characterize whether the planet transits or not: 
the PA must be constant for a transiting planet. In the position measurements of MB15 
the PA is found to be higher than in the previous studies, which leads to the conclusion 
that the inclination of the orbit is significantly different from 90$\degr$.

\begin{figure}[tb]
\includegraphics[angle=90,width=\columnwidth]
{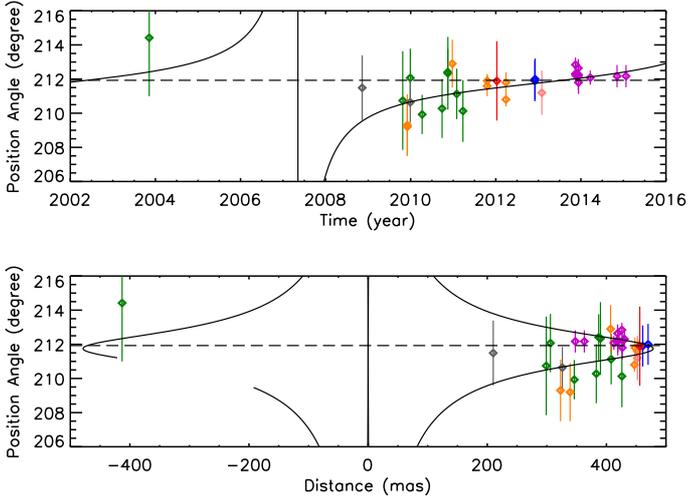}
\caption[]{
Plot of the measured position angle of \bpicb\ as a function of time 
and as a function of the distance to the star.
Vertical bars represent 1-$\sigma$ error bars. 
The colors of the symbols are the same as in Fig.~\ref{Position_Envelop_Fig7}. 
The gray point is the single measurement of Currie et al.\ (2011), which was obtained 
very early at a low projected distance; given its uncertainty, it does not discriminate between the two scenarios, unlike the measurements of MB15 (purple). 
The model with an orbit inclination of 89$\degr$ (solid line) fits the data more closely
than an inclination of 90$\degr$ (constant value of PA obtained by weighted mean of the data, dashed line).
\label{PA_vs_Time}}
\end{figure}

We plotted and fitted the PA measurements as a function of time and as a function of the distance to the star
(Fig.~\ref{PA_vs_Time}). The model with an orbit inclination of 89$\degr$ yields a better fit to the data 
($\chi^2$=19.2 for 32~degrees of freedom) than the model with an orbit inclination of 90$\degr$ 
($\chi^2$=29.7). Hence, in agreement with MB15, we find a 3-$\sigma$ significance 
of a non-transiting orbit ($\sqrt{\Delta\chi^2}$=$\sqrt{10.5}$=3.2).
Therefore, there is a first scenario, where the planet does not transit. Nonetheless, as noted by MB15, 
in this scenario the Hill sphere of the planet still does transit the star; 
with a rich environment as suspected in the case of J1407b (Kenworthy \& Mamajek 2015) 
the transit of the Hill sphere could also be responsible for the photometric event seen in 1981.

There is, however, a second scenario: MB15 measurements of the PA can have a systematic offset 
of about 0.5$\degr$ relative to the other measurements. 
We can indeed imagine the possibility that different teams yield different 
absolute calibration of the PA (see fifth paragraph of Sect.~1 in MB15). 
We do not judge that one is better than another, but we acknowledge that the calibration 
of the PA in absolute value is very difficult with many possible systematic effects 
(see detailed description of the process in Sect.~4.1 of MB15). 
Consequently, there could be a systematic difference of 0.5 degrees depending 
on the calibration process used by different teams. 
To illustrate this possibility, we note that among the ten~measurements published by MB15, 
two are obtained on images that had already been analyzed by others (Macintosh et al.\ 2014  
and Bonnefoy et al.\ 2014). From the same data sets, 
these authors found a PA lower by 0.42$\degr$ and 0.56$\degr$, respectively. 
Indeed, if we decrease the PA measurements of MB15 by only 0.5$\degr$, 
then we obtain the same $\chi^2$ for a 90$\degr$ inclination orbit ($\chi^2$=19.5) 
as for the 89$\degr$ inclination orbit using the original MB15 measurements. 
Similarly, if we do not take the measurements of MB15, but rather the values published 
by Macintosh et al.\ (2014) and Bonnefoy et al.\ (2014), then the model with an 89$\degr$ 
inclination is not significantly better than the model with a 90$\degr$ inclination.

In summary, the orbit of \bpicb\ is very close to the transit configuration with a very low 
angle between the orbital plane and the line of sight; this angle is significantly lower 
than the angle of the disk warp (seen in projection), which is about 4$\degr$ 
(Apai et al.\ 2015). 
As a consequence, it is certain that the Hill sphere of \bpicb\ transits the star. 
It is also possible that the planet itself effectively transits, if the latest PA measurements 
of MB15 have a systematic shift of about 0.5$\degr$ relative to all previous publications due to 
a different calibration process.

As seen in Fig.~\ref{PA_vs_Time}, new measurements to be taken in 2016 will allow us to
discriminate between the two scenarios. 

\section{New fits to the orbit}

In the following, we assume that the imaged planet is the same 
as the transiting object in November 1981. 
With this assumption, we obtain new constraints on the possible 
orbital characteristics of \bpicb\ (Sect.~\ref{Results}), 
and we make predictions for the forthcoming observations 
(Sect.~\ref{Discussion}). Comparison of these predictions 
with observations will allow us to disprove or endorse the present model. 
First we describe the available data (Sect.~\ref{data}), and then the method 
used to constrain the orbit of \bpicb\ (\ref{method}).

\subsection{Data}
\label{data}

The data to be fitted are the star-planet distance measurements, 
the radial velocity of the planet, and the estimate of the stellar mass used 
for the relationship between the semi-major axis and the orbital period.

The planet position measured on high resolution images are tabulated 
in various papers listed in the caption of Fig.~\ref{Position_Envelop_Fig7}. 
Here, because we assume that the planet is a transiting planet, 
we used only the measurements of the star-planet distance to constrain the orbit. 
Checking the error bars estimates, we saw that the error bars 
of Nielsen et al.\ (2014) are likely underestimated by a factor of $\sim$2. 
A fit of these measurements shows that the dispersion, 
given by the $rms$, is about twice the error bars. 
This dispersion larger than the tabulated error bars 
is also seen by comparing the four measurements made in October 2011 and March 2012: 
there is a decrease of 6~milliarcsec between these two epochs with error bars 
of 3 or 5~milliarsec, while an increase 
of 12~milliarcsec is expected to be measured. As a consequence, 
all the fits using this data set yield values of the $\chi^2$ 
that are significantly higher than the number of degree of freedom, or, equivalently, 
a reduced $\chi^2$ significantly larger than~1 ({e.g.}, MB15). 
Therefore, we decided to multiply by a factor
of~2 the error bars on the star-planet distance tabulated by Nielsen et al.\ (2014).

For the two published estimates of the star-planet distance obtained from the same observations 
(Macintosh et al.\ 2014; Bonnefoy et al.\ 2014; and MB15, see Sect.\ref{Orbit inclination}), 
we used the weighted mean of the two values. 

An important piece of information used to constrain the orbit of \bpicb\ is the measured radial velocity of the planet. 
Using VLT high contrast and high resolution spectroscopic observations in the infrared, Snellen et al.\ (2014) 
provided a measurement of the planet radial velocity of -15.4$\pm$1.7\,km/s in December 2013. This measurement
shows that the planet orbits in the same direction as the gas disk (Olofsson et al.\ 2001; 
Brandeker et al.\ 2004) and that the planet motion between 2009 and 2015 is seen toward the observer, 
{i.e., } at the next conjunction the planet will be in front of the star. 
We included this measurement in all our fits to constrain the planet orbit.

We assume that the \bp\ stellar mass is 1.75$\pm$0.05\,$M_{\odot}$ (Crifo et al.\ 1997).
The distance to $\beta$\,Pic is given by Hipparcos measurements as $d$=19.3$\pm$0.2\,pc (Crifo et al.\ 1997).
This distance is used to translate the angle distance between the star and the planet into 
a projected distance from the star in astronomical units. 
The stellar mass is considered in the fit by making it a free parameter. 
The stellar mass estimates provided by the stellar physics (Crifo et al.\ 1997) 
can hence be included in the fit procedure by adding its value in the calculation of the $\chi^2$. 
We also fitted the data ignoring that information, which provides an independent estimate 
of the \bp\ stellar mass by only the observations of its planet orbit (Sect. \ref{Results}). 

\subsection{Method}
\label{method}

We fitted the data (star-planet distances and the radial velocity of the planet)
using four free parameters: three orbital parameters and the stellar mass $M_s$. 
Three parameters are enough to describe the orbit because the planet 
is assumed to be on an edge-on orbit and to have transited on November 10, 1981. 
The inclination is thus fixed to 90$\degr$ (removing a parameter for the longitude of the node), 
and the epoch of the periastron is constrained by the date of the transit.
The free orbital parameters are 
the following: the orbital period $P$, $e\sin \varpi $, and $e\cos \varpi$, 
where $e$ is the orbital eccentricity and $\varpi$ the argument of periastron. 

To search for the best fit, we first estimate the parameters 
by running a Levenberg-Marquardt $\chi^2$ minimization algorithm.
The uncertainties are then estimated using a Metropolis-Hasting Markov chain 
Monte Carlo (MCMC) algorithm ({e.g.}, Tegmark et al.\ 2004) 
with an adaptive step size. The parameter space is mapped 
with a total of $5\times 10^7$ steps in the chains 
(see Bourrier et al.\ 2015 for more details on our MCMC). 

\section{Results}
\label{Results}

\begin{table}[tbh]
\caption{Orbital parameters of \bpicb\ assuming a transit in November 1981. 
Error bars are 1-$\sigma$ intervals corresponding to 68\% of the posterior distributions, 
except for the epoch of the next transit given with 2-$\sigma$ intervals. 
The $\chi^2$ are for 30 degrees of freedom (34 measurements and 4 free parameters).} % title of Table
\label{Table:1} % is used to refer this table in the text
\centering % used for centering table
\begin{tabular}{c c c c} % centered columns (4 columns)
\hline\hline % inserts double horizontal lines
\noalign{\smallskip}
Parameter & Low          & High         & Units\\ % table heading
          & eccentricity & eccentricity & \\ % table heading
\hline % inserts single horizontal line
\noalign{\smallskip}
$\chi^2$                 & 36.5              & 23.8 \\
\hline % inserts single horizontal line
\noalign{\smallskip}
Period $P$                                               & 17.97 $\pm$ 0.08  & 36.38 $\pm$ 0.13 & years \\
Semi-major\\
axis $a$ & 8.20 $\pm$ 0.06  & 13.18 $\pm$ 0.09 & au \\
Eccentricity $e$       & 0.118 $\pm$ 0.020 & 0.323 $\pm$ 0.005& \\               
Longitude of \\ 
periastron $\varpi$    & -108 $\pm$ 28      & 87.0$\pm$5.6 & degrees\\ 
$e \cos \varpi$                          & -0.04 $\pm$ 0.06  & -0.04 $\pm$ 0.03 & \\  
$e \sin \varpi$                          & -0.097$\pm$ 0.006 & 0.319 $\pm$ 0.004 & \\
Star mass $M_s$        & 1.71 $\pm$ 0.04   & 1.73 $\pm$ 0.04 &  $M_\odot$ \\
\hline % inserts single horizontal line
\noalign{\smallskip}
Range of epochs           &  15/Jul/2017 &   1/Jan/2018 & \\
for the                   &     -       &    -         & \\
next transit (2-$\sigma$) & 1/Mar/2018 &  30/Jun/2018 & \\
\hline %inserts single line
\end{tabular}
\end{table}

\begin{figure*}[p!]
\begin{center}
\includegraphics[angle=90,width=0.85\textwidth]
{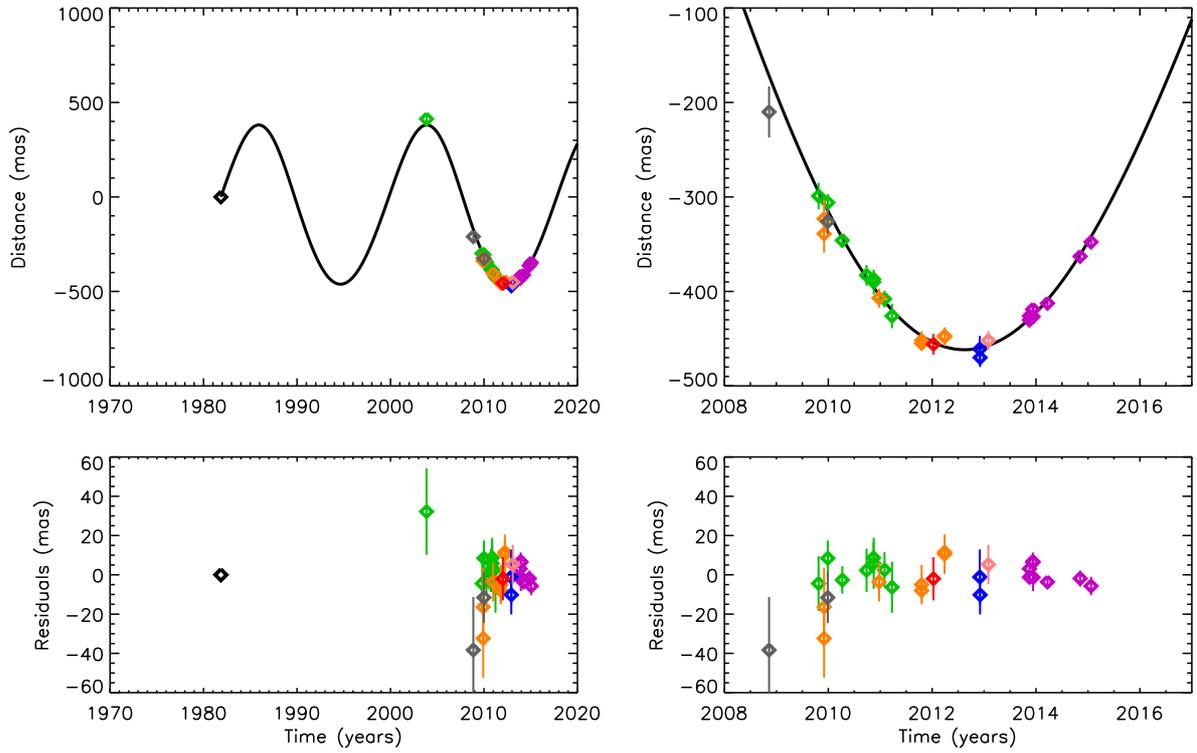}
\end{center}
\caption[]{ Projected distance between \bpicb\ and its host star in milliarcsecond (mas) as a function of time. 
The colors of the symbols are the same as in Fig.~\ref{Position_Envelop_Fig7}. 
The upper panels show the measured positions of the planet with their error bars, along with the best fit obtained 
with an orbital period of about 18.0~years. The bottom panels show the
residual differences between the measurements and the fit with the assumption that the planet transited in 1981. 
The right panels show a zoom on the projected distances and residuals for the 2009-2015 time period. 
\label{Orbit_circ}}
\end{figure*}

\begin{figure*}[p!]
\begin{center}
\includegraphics[angle=90,width=0.85\textwidth]
{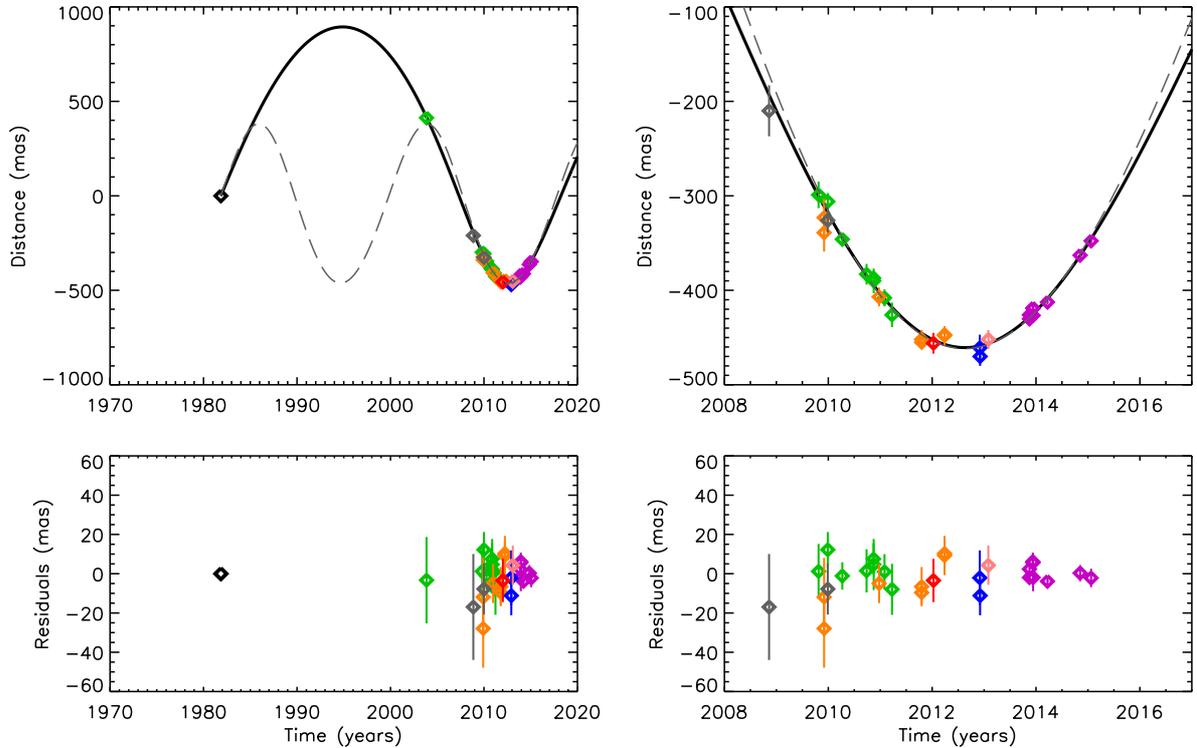}
\end{center}
\caption[]{Same as in Fig.~\ref{Orbit_circ} for the solution with the long orbital period of $\sim$ 36.4~years.
The fit with the orbital period of 18~years is overplotted with a gray dashed line. 
\label{Orbit_ecc}}
\end{figure*}

We identified two minima of the $\chi^2$ in the parameter space. 
These minima are deep enough such that the MCMC chains are always trapped 
in one of them and never escape from it. 
The resulting orbits are shown in Figs.~\ref{Orbit_circ} and~\ref{Orbit_ecc} and the corresponding
parameters are given in Table~\ref{Table:1}.

There are two families of orbits that are consistent with the observations. 
The first has a low eccentricity of $\sim$0.1 and an orbital period of about 18~years.
The second family of orbits yields a larger orbital period around 36.4~years
and a higher eccentricity above 0.3.

\subsection{The orbit with an 18-year period}

\begin{figure*}[p!]
\begin{center}
\includegraphics[angle=90,width=0.8\textwidth]
{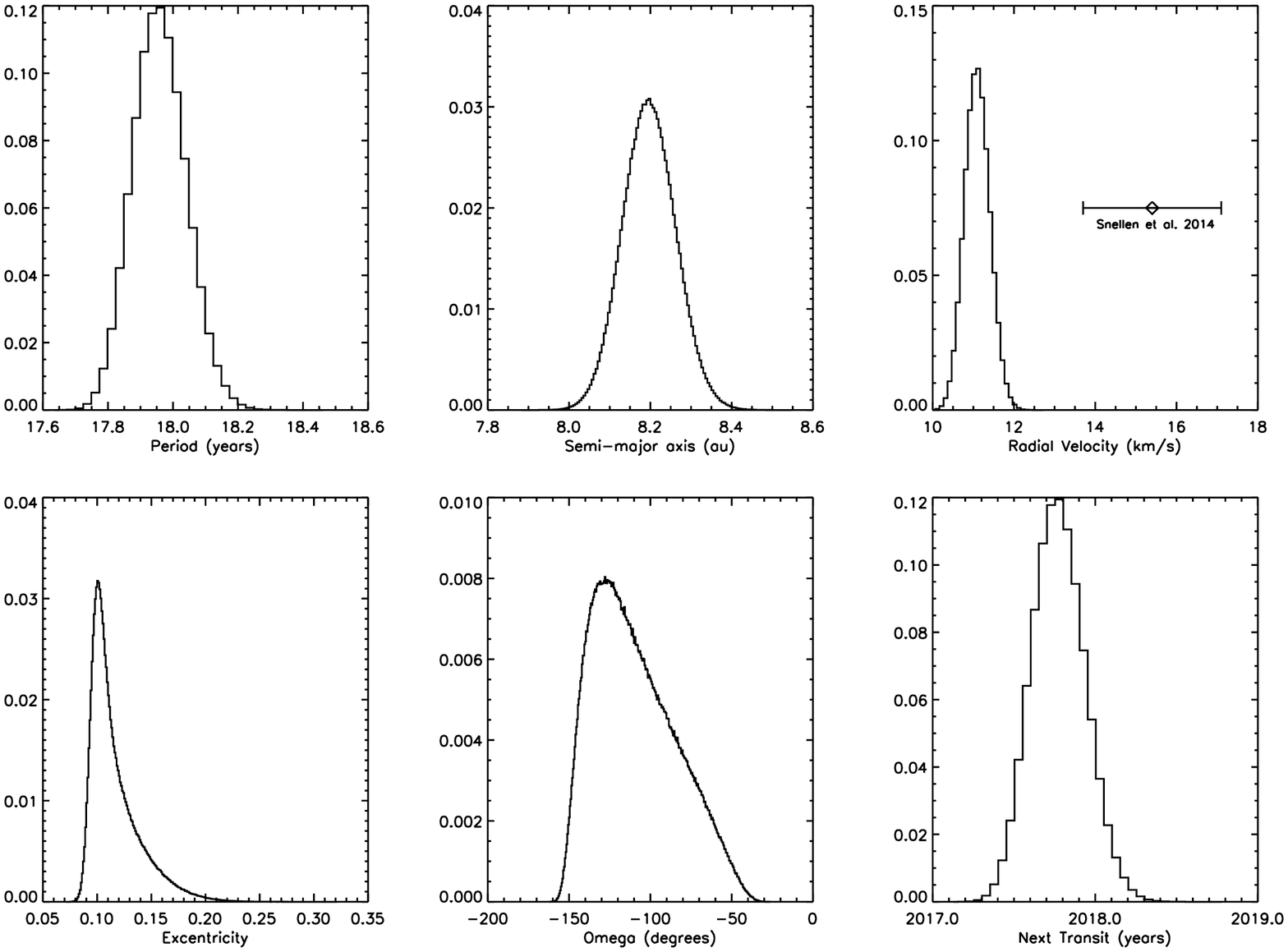}
\end{center}
\caption[]{Plot of the distribution of the orbital period, semi-major axis, eccentricity, 
longitude of the periastron ($\varpi$), and time of the next transit for the solutions 
with an orbital period around 18.0~years. The radial velocity of \bpicb\ at the epoch 
of the measurement of Snellen et al.\ (2014) is plotted in the top right panel. 
\label{MultiPlot_circ}}
\end{figure*}

The first family of orbits is found when the starting point 
in the parameter space is at low eccentricity. 
In this case we find an orbit with an orbital period of 18~years, in which
\bpicb\ also transited in front of the star in 1999-2000 
(Fig.~\ref{Orbit_circ}). 
This orbit is similar to the one found in previous works 
(Bonnefoy et al.\ 2014; Nielsen et al.\ 2014; MB15).
The posterior distributions of the orbital period, semi-major axis, eccentricity, 
longitude of the periastron ($\varpi$), and time of the next transit 
are given in Fig.~\ref{MultiPlot_circ}. 

In this case the $\chi^2$ is significantly larger than the number of degrees of freedom.
This is in agreement with 
Nielsen et al.\ (2014) and MB15, who also found a reduced $\chi^2$ significantly above 1. 

Moreover, if we do not add the estimate of the mass of the star as given 
by the stellar physics into the $\chi^2$ calculation, 
we then obtain a stellar mass constrained only by the orbital motion of
the planet. In this case we find a stellar mass of 1.64$M_\odot$$\pm$0.06$M_\odot$, i.e.,
 about 2-$\sigma$ lower than the estimate of Crifo et al.\ (1997). 
Millar-Blanchaer et al. (2015) reached the same conclusion. 

Finally, although the measurement of the radial velocity of the planet (Snellen et al.\ 2014)
is included in the fit to the data, the low eccentricity orbit yields a radial 
velocity at the epoch of the measurement that is 2.5-$\sigma$ lower 
than the measured one (Fig.~\ref{MultiPlot_circ}); 
MB15 also faced the same puzzling result. 

In conclusion, a low eccentricity orbit 
is a possible solution to the available measurements of \bpicb .
This solution is the only one considered in the published paper on the \bpicb\ orbit. 
However, this solution does not provide a very good fit to the data. 
In summary, the high $\chi^2$ for the low eccentricity orbit is caused by a 
bad fit to the following data: the radial velocity of the planet, the stellar mass, 
and also the early measurements of Currie et al.\ 2011 (not included in the analysis of MB15). 
These are three different and independent physical pieces of information on the system; 
they might have been under- or overestimated, 
or their error bars might possibly have been underestimated. 
Nonetheless, in this context, the existence of another family of solutions with a higher
eccentricity, which provides a better fit to the data, is to be considered with interest. 
 
\subsection{The orbit with a 36-year period}

In our initial study with only the 2003 observation available, 
we also found a second possible orbital solution with a semi-major axis of about 17~AU 
and an orbital period of about 52~years 
(see Fig.~6 of Lecavelier des Etangs \& Vidal-Madjar 2009). 
In this case, the planet had moved along a little less than half of an
orbit from 1981 to 2003, and was observed in 2003 at a projected distance of 8~AU just before 
the opposition (or secondary transit).
In this solution, the planet was at a greater projected distance from \bp\ before 2003. For instance, 
in 1995, the planet would have been at 0.9 arcsec from \bp\ and might have been detected in HST 
or adaptative optics ground-based observations; thus in our initial study we concluded that this solution 
was less likely than the solution with an orbtial period of 18~years. 

\begin{figure*}[p!]
\begin{center}
\includegraphics[angle=90,width=0.8\textwidth]
{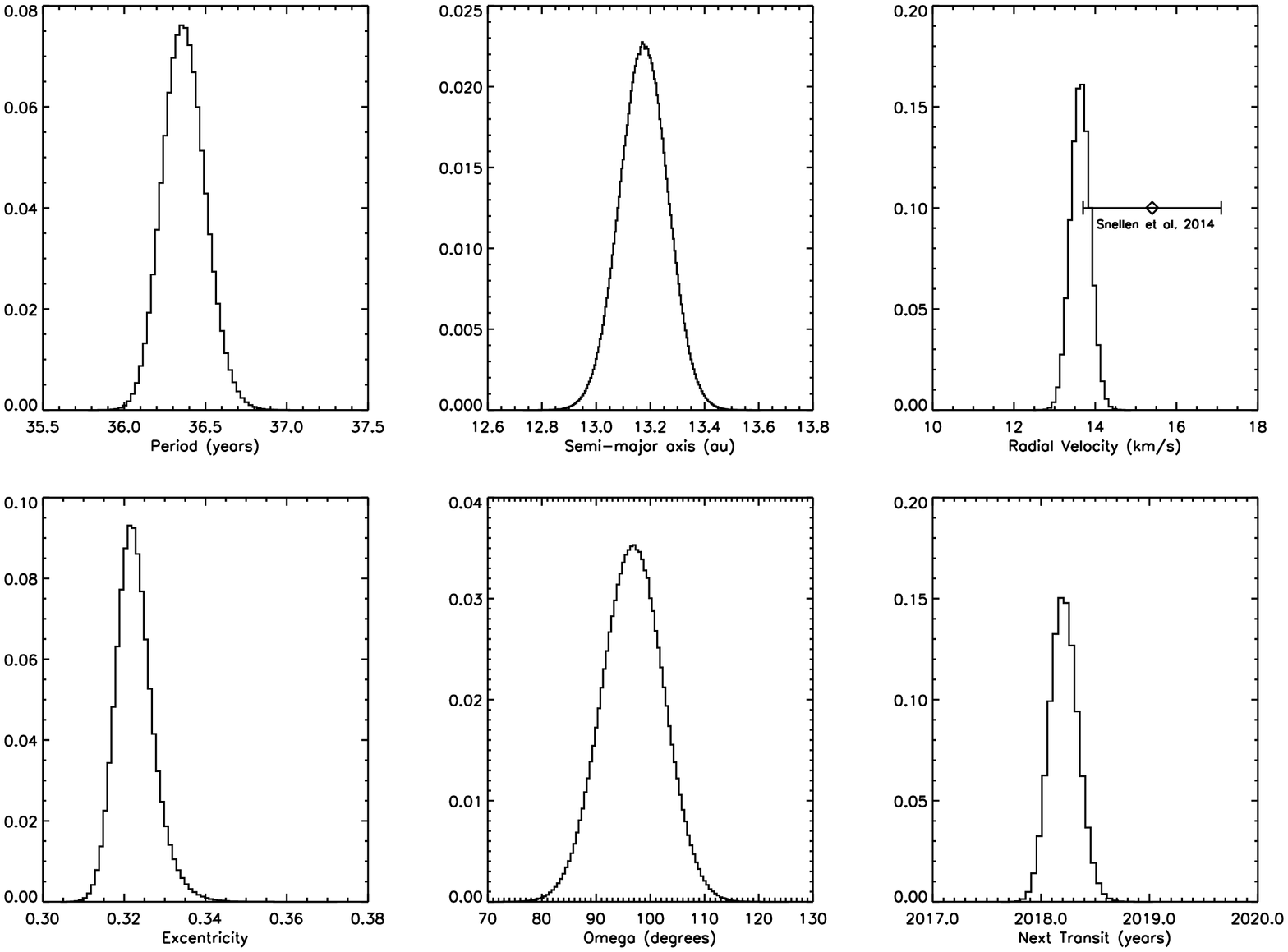}
\end{center}
\caption[]{Same as in Fig.~\ref{MultiPlot_circ} for the solution 
with the long orbital period of $\sim$ 36.4~years. 
\label{MultiPlot_ecc}}
\end{figure*}

However, with all the new data available now, a second deep minimum of $\chi^2$ 
appears in the parameter space. 
We find an orbit with an orbital period of 36.4~years, in which
\bpicb\ has not transited in front of the star since 1981 (Fig.~\ref{Orbit_ecc}). 
In this case the fit is much better than the fit with a low eccentric orbit, 
with a $\chi^2$ that is lower than the number of degrees of freedom
(Table~\ref{Table:1}).
The corresponding posterior distributions of the orbital parameters,
the time of the next transit, and the radial velocity of the planet 
are given in Fig.~\ref{MultiPlot_ecc}. 

Moreover, using the high eccentric orbit, we obtain a stellar mass constrained 
only by the orbital motion of the planet, which is now consistent at 1-$\sigma$ 
with the stellar physics: 1.69$M_\odot$$\pm$0.06$M_\odot$. 

In addition, as shown in Fig.~\ref{MultiPlot_ecc}, the high eccentricity orbit 
yields a radial velocity of the planet that is consistent with the velocity 
of \bpicb\ measured by Snellen et al.\ (2014).
We reached the same conclusion even if we do not include this measurement 
in the $\chi^2$ calculation of the fit: the radial velocity of \bpicb\ 
measured in December 2013 points toward a high eccentricity orbit. 

In conclusion, a high eccentricity orbit is consistent with all the available measurements of \bpicb . 
This provides a much better fit to the measurements of the planet's astrometric position
(including the early measurements of Currie et al.\ 2011), the planet radial velocity, 
and the estimates of the mass of the star provided by the stellar physics. 

\section{Discussion and future observations}
\label{Discussion}

\subsection{Discussion}

We have found a solution for the \bpicb\ orbit with a 36.4~years orbital period, which did not 
show up in the previous studies ({e.g.}, Chauvin et al.\ 2012; MB15). 
We suspect that this is because these previous studies searched for solutions 
in a broader parameter space (not imposing the orbit to be seen exactly edge-on as for a transiting planet) 
and because the addition of a transit in 1981 put a strong constraint on the planet orbit and significantly 
modify the geometry of the possible solutions in the parameter space.
 
In both cases the solutions are still plausible, but the solution 
with a 36.4-year orbital period better fits the available data with the hypothesis of a transit in 1981. 
The forthcoming observations in 2016 (or later) should allow us to 
discriminate between these two types of orbit because the two orbits diverge after 2015
(Fig.~\ref{Orbit_ecc}).

\subsection{Schedule of the forthcoming transit}

For extrasolar planets studies, the transit is a key configuration
for characterizing the orbit 
({e.g.}, H\'ebrard \& Lecavelier des Etangs 2006), 
to search for the photometric signature of evaporating bodies 
({e.g.}, Lecavelier des Etangs et al.\ 1999),
or to search for atmospheric signatures 
({e.g.}, Vidal-Madjar et al.\ 2003). 

If \bpicb\ is a transiting planet, it would be of prime interest to know 
when the next transit will happen. 
Assuming that the planet has an orbital period of 18~years, 
we find that the next primary transit should happen 
between 15$^{}$ July 2017 and 1$^{}$ March 2018 
(2-$\sigma$ confidence interval). 
With an orbital period of 36.4~years, 
the next primary transit should happen between 
1$^{}$ January 2018 and 30$^{}$ June 2018 
(2-$\sigma$ confidence interval). 
Forthcoming 2016 observations should help to better predict this extremely important event.
Even if this primary transit is preceded by light variations over a few days (Lecavelier et al.\ 1995), it 
is extremely difficult to anticipate this kind of observation. It is more likely that new image
observations in 2016 will help to better constrain the time of the next transit. 

\section{Conclusion}

We have constrained the possibilities for future observations
of \bpicb\ assuming that it was the transiting object of 1981. 
After acknowledging the uncommon potentialities of this transiting planet, 
we are still waiting for new data to improve or refute the proposed scenario. 

Presently, continuing or starting new photometric surveys of \bp\ will be useful. 
For example, long timescale photometric variations due to the occulting belt 
of dust in 1:1 resonance with the planet (Lecavelier des Etangs et al.\ 1997) 
could be used to give a warning of the next transit event. 
Indeed, dust accumulated close to the Lagrange point should be responsible for
some variations in the extinction when it moves in front of the star. 
The highest extinction should be reached about 3 years before the transit, 
i.e., in $\sim$2015, with a decrease in the extinction later on. 
If these long timescale photometric variations are detected, they could allow us 
to anticipate the next transit and to confirm
the link between the photometric event of 1981 and \bpicb .

If future observations happen to confirm that \bpicb\ is a transiting planet, 
this planet would be an extraordinary transiting planet, because it is 
\begin{enumerate}
\item a planet transiting in front of a 4$^{\rm th}$ magnitude star. 
By comparison, with what has been done in the case of planets transiting 
7$^{\rm th}$ magnitude stars ({e.g.}, Charbonneau et al. 2002; Sing et al.\ 2008), 
the atmosphere of this planet could be probed with unprecedented details 
for an extrasolar planet; 
\item a planet with a very long transit duration of several hours. 
This could allow further improved studies of the planetary atmosphere 
including structures of the atmosphere along the planetary limb;
\item a young planet with circumplanetary material. 
Detailed transit observations could give unique information on the planet environment 
including rings and satellites at a stage when satellites are still forming or just formed;
\item a planet with a semi-major axis of 8 or 13~au. This would give access to transit observations 
of a planet far from its parent star, a situation more like the giant planet of the solar system, 
and much different from the short period exoplanets, presently the only known transiting planets. 

\end{enumerate}
 
In conclusion, we are now waiting for new observations to confirm or disprove 
the hypothesis developed in this present work. If confirmed, \bpicb\ could soon become 
a mine of information on young extrasolar planets. 

\begin{acknowledgements}
We warmly thank C.~Nitschelm who first pointed out the photometric measurements of \bp\ obtained 
by the Geneva Observatory. 
We are also grateful to R.~Ferlet, G.~H\'ebrard, and P.A.~Wilson for fruitful discussions 
on the subject of the present work. This work has been supported
by an award from the Fondation Simone et Cino Del Duca.
We acknowledge the support of the French Agence Nationale
de la Recherche (ANR), under program ANR-12-BS05-0012 ``Exo-Atmos''. 

\end{acknowledgements}

\end{document}